%% file: main.tex
\documentclass[a4paper,fleqn]{cas-sc}
\usepackage{adjustbox}
\usepackage{float}
\usepackage[numbers]{natbib}
\def\tsc#1{\csdef{#1}{\textsc{\lowercase{#1}}\xspace}}
\tsc{WGM}
\tsc{QE}
\tsc{EP}
\tsc{PMS}
\tsc{BEC}
\tsc{DE}
\begin{document}
\let\WriteBookmarks\relax
\def\floatpagepagefraction{1}
\def\textpagefraction{.001}
\shortauthors{S.Mu et~al.}

\shorttitle{IMC-PINN-FE}    

\title [mode = title]{IMC-PINN-FE: A Physics-Informed Neural Network for Patient-Specific Left Ventricular Finite Element Modeling with Image Motion Consistency and Biomechanical Parameter Estimation}  

\author[1]{Siyu Mu}[]
\ead{siyu.mu21@imperial.ac.uk}

\affiliation[1]{organization={Department of Bioengineering, Imperial College London},
                state={London},
                country={UK}}

\author[1]{Wei Xuan Chan}[]

\author[1]{Choon Hwai Yap}[orcid=0000-0003-2918-3077]
\ead{c.yap@imperial.ac.uk}

\cortext[1]{Corresponding author:}

\begin{abstract}
Elucidating the biomechanical behavior of the myocardium is important for understanding of cardiac physiology, but this cannot be directly inferred from clinical imaging and requires finite element (FE) simulations. However, conventional FE simulations are computationally expensive and often do not accurately reproduce clinically observed cardiac motions. To address these limitations, we propose the IMC-PINN-FE, a physics-informed neural network (PINN) framework that integrates imaged motion consistency (IMC) with FE modeling for patient-specific simulations of left ventricular (LV) biomechanics. Cardiac motion is first estimated from 4D MRI or echocardiography using either a pre-trained attention-based network or an unsupervised cyclic-regularized network, followed by extraction of motion modes. IMC-PINN-FE then rapidly estimates myocardial stiffness and active tension by fitting end-diastolic (ED) and peak-systolic (PS) clinical measurements. As an unsupervised PINN-based estimator, IMC-PINN-FE significantly accelerates biomechanical parameters estimation, reducing computation times from several hours to mere seconds compared to conventional inverse FE methods. Finally, based on these computed parameters, IMC-PINN-FE performs FE modeling throughout the cardiac cycle, enabling computational speeds 75 times faster than traditional FE simulations. Through motion constraints, its output of cardiac displacements across the cardiac cycle matches imaged motions much more accurately, achieving an average Dice coefficient of 0.927, which is markedly higher than 0.849 for conventional FE methods. At the same time, it preserves a realistic pressure–volume relationship that remains very similar to that of conventional FE. Unlike existing PINN models, IMC-PINN-FE introduces explicit motion fidelity constraints and enables back-computation of myocardial properties. Our strategy of using cardiac motions of a single individual to reconstruct shape mode avoids the need for a large dataset, and enables better patient-specificity. IMC-PINN-FE thus presents a robust approach for efficient, patient-specific, and image-consistent cardiac biomechanical modeling. Code is available at \url{https://github.com/SiyuMU/IMC_PINN_FE} and \url{https://github.com/SiyuMU/EFDL}.

\end{abstract}

% Keywords

\maketitle

% Main text
\section{Introduction}
\input{Introduction}
\section{Methods}
\input{Methods}

\section{Results}
\input{Results}
\section{Discussion and Conclusion}
\input{Discussion}

\section*{Acknowledgment}
This work was supported by a PhD studentship funded by Imperial College London. Additional support was provided by the Additional Ventures Single Ventricle Research Fund \#1019496.

\bibliographystyle{cas-model2-names}

\bibliography{cas-refs}

\end{document}

%% file: Introduction.tex
Computation of the biomechanical behavior of the heart is useful for evaluating cardiovascular diseases and can facilitate the development of cardiac digital twins for optimized and personalized medicine. Cardiac biomechanics is traditionally computed through image-based finite element (FE) simulations. This has previously been performed for both healthy and pathological cardiac conditions \cite{wang2015image, green2024myocardial}, providing details of myocardial deformation, stress distribution, and contractile function \cite{hurtado2015improving}, which offered physiology insights into diseases and intervention scenarios \cite{green2024pre}.\\

However, current FE-based approaches face significant challenges in clinical applications. Traditional cardiac FE methods rely on time-consuming simulations, making real-time, patient-specific modeling impractical \cite{zhang2020novel}. Additionally, most FE models assume population-averaged myocardial properties, such as stiffness and active tension, and are thus not fully patient-specific \cite{shavik2018high}. Moreover, existing FE methods often struggle to faithfully replicate the complex, time-dependent myocardial deformations observed in MRI and ultrasound data \cite{salih2024advancements}. While iterative tuning of the FE model to back-compute myocardial properties and to better match clinically measured motion is possible \cite{genet2018equilibrated}, it is both time- and resource-intensive. Therefore, FE models have not yet been widely adopted in clinical practice. This highlights the need for an efficient, patient-specific, and physiologically consistent alternative, one that is fast enough to be clinically adoptable, and that integrates image-based motion tracking with physics-informed biomechanical constraints.\\

Recent advances in physics-informed neural networks (PINNs) offer a promising approach to achieve this, combining deep learning (DL) with physical constraints to bypass the need for explicit iterative solvers \cite{raissi2019physics}. Several PINN-FE approaches have recently been proposed for modeling left ventricular (LV) biomechanics. Motiwale et al. developed a PINN-FE framework that predicts the nodal displacements of an idealized 3D LV mesh from pressure and active tension inputs, guided by the governing equations in their weak form \cite{motiwale2024neural}. Buoso et al., in contrast, incorporated a pre-computed statistical shape model to encode cardiac geometry and motion, enabling a similar input–output mapping across a variety of realistic cardiac anatomies \cite{buoso2021personalising}.  Their use of shape encoding significantly reduced computational costs. However, these methods do not enforce an adherence to imaged myocardial motion, which limits their adaptability to patient-specific variations. Moreover, they lack a mechanism for inverse estimation of myocardial stiffness and active contractile tension.\\

To address these challenges, we propose Image Motion Consistent PINN-FE (IMC-PINN-FE), a novel framework that integrates patient-specific motion tracking into a reduced-order PINN-FE model. IMC-PINN-FE encodes cardiac motion using modes extracted from cine-imaging data of an individual subject rather than from a large dataset of diverse individuals. This helps to retain the computational efficiency of reduced-order modeling while eliminating the requirement for an extensive dataset, and enhances the patient-specificity of modeled motions. Furthermore, the framework is constrained to match observed LV volumes from clinical images, and it incorporates inverse estimation of myocardial stiffness and peak active tension prior to performing detailed biomechanical simulations.\\

The framework assumes the availability of accurate 3D reconstructions of the LV geometry as input. In practice, several fast and reliable tools can reconstruct 3D geometries directly from clinical images, including 3D segmentation algorithms \cite{wiputra2016fluid, ilesanmi2024reviewing} and 3D mesh reconstruction methods \cite{van20183d, luo2024explicitdifferentiableslicingglobal}. Next, the framework requires accurate cardiac motion tracking from the images. For this, we propose two options: a novel unsupervised network that enforces cyclic consistency in pairwise image registration, and a previously reported cross-attention-based deep learning model \cite{ahn2023co}. These tools provide the necessary anatomical and motion inputs to support personalized modeling in IMC-PINN-FE.\\

The key contributions of this work are as follows:
\begin{enumerate}
\item \textbf{Motion-Consistent PINN-FE}: We propose the IMC-PINN-FE, a framework for rapid tissue biomechanics modeling of the cardiac ventricle that enforces consistency with both physics-based governing equations and 4D image-derived cardiac motion, thereby improving the physiological accuracy of biomechanical assessments. This represents a departure from prior approaches that primarily focused on enforcing physics constraints. Motion consistency is achieved through the use of patient-specific cardiac motion modes and constraints based on imaged ventricular volumes. 

\item \textbf{High Computational Efficiency}: IMC-PINN-FE leverages image-derived motion modes for reduced-order modeling, enabling a reduction in computational time by approximately 75× compared to traditional FE methods, while maintaining fidelity to patient-specific motion. This efficiency makes the approach feasible for real-time or near real-time clinical applications. Further, our use of cine-images of a single patient rather than a large multi-individual dataset for statistical shape modelling \cite{buoso2021personalising} is less cumbersome and more subject-specific.

\item \textbf{Efficient Physics-Informed Parameter Estimation}: IMC-PINN-FE performs fast, unsupervised estimation of myocardial stiffness and peak active tension, enhancing personalization and supporting accurate patient-specific biomechanical modeling.
\end{enumerate}

%% file: Methods.tex
\begin{figure}
	\centering
	\includegraphics[width=\textwidth]{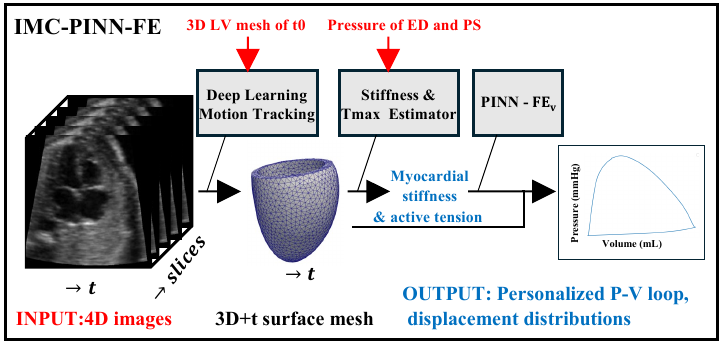}
	\caption{The overall framework of IMC-PINN-FE: inputs to the framework include 4D (3D+time) imaging data (e.g., echocardiography or MRI) combined with LV reconstruction obtained via deep learning (DL) and physiological inputs, including peak systolic (PS) and end-diastolic (ED) LV pressures. Outputs include estimates of patient-specific LV myocardial stiffness and maximum active tension, and LV finite element (FE) simulation results across the cardiac cycle.}
	\label{FIG:schematic_overall}
\end{figure}

\def\displacement{d\overrightarrow{X}_{t \rightarrow{} t+\delta t}}

The framework of IMC-PINN-FE, as illustrated in Fig. \ref{FIG:schematic_overall}, follows a three-step pipeline that integrates deep learning and physics-informed principles: 

\begin{enumerate}
\item \textbf{Deep Learning (DL) Cardiac Motion Tracking}: A DL image registration (DLIR) network is employed to extract smooth and cyclic cardiac motion from 4D medical images. Two approach options are proposed: (a) a novel network that ingests rapidly computed pairwise 3D registrations and outputs a 4D cyclic motion field parameterized  by temporal Fourier series and spatial B-splines; and (b) a previously reported deep learning cross-attention network \cite{ahn2023co}. Approach (a) does not require pre-training over a large dataset, but takes more time than approach (b). 

\item \textbf{Back-Computing Myocardial Stiffness and Active Tension}: Using the estimated motion and inputted LV pressure information, a PINN network predicts myocardial stiffness and maximum active tension, based on FE governing equations. 

\item \textbf{PINN-FE solver}: This final step uses the same solver as in (2) to perform FE modeling for the whole cardiac cycle, where shape mode coefficients describing displacements are obtained under the constraints of the governing equations and imaged LV volumes guidance.
\end{enumerate}

\subsection{Deep learning Motion Tracking}
Mainstream 4D (3D+t) image registration techniques can be broadly categorized into two types: classical non-DL methods, such as B-spline \cite{heydeMotionDeformationEstimation2012, chan2021full} and optical flow \cite{horn1981determining} registration, and DL methods, including co-attention spatial transformer networks (STN) \cite{ahn2023co} and feedback attention DLIR \cite{10843372}.\\

While non-DL methods do not require large datasets for training, they are often computationally intensive and time-consuming, particularly for 4D image registration tasks. In contrast,  DL approaches offer significantly faster inference but depend heavily on large training datasets, posing challenges due to the labor-intensive process of manually labeling segmentation ground truth in 4D and the substantial memory demands associated with 4D networks. These difficulties become even more pronounced given that DL models frequently require retraining when adapted to a different imaging modality or imaging condition (e.g., transitioning from fetal to adult echocardiography, or from echocardiography to MRI).\\

Given this trade-off, we systematically compared several representative methods to identify the most suitable candidate for integration to IMC-PINN-FE. We also introduce a novel unsupervised DL cardiac motion estimation algorithm, termed Elastix Fourier Deep Learning (EFDL). EFDL performs a spatial B-spline of temporal fourier cyclic regularization on computationally efficient pairwise registration via a network. This hybrid approach combines the strengths of both traditional and learning-based methods: it eliminates the need for large-scale pretraining datasets, while achieving faster computation compared to its non-DL counterpart, the B-spline Fourier method \cite{chan2021full, wiputra2020cardiac}.\\

\subsubsection{Elastix Fourier Deep Learning (EFDL)}
The EFDL framework (detailed schematic in supplementary S1) consists of a rapid pairwise registration followed by Fourier regularization.

\paragraph{\textbf{Bspline Pairwise Registration}:} 
During the rapid registration stage, any pairwise registration algorithm may be employed. In this study, we adopt the cubic B-spline free-form deformation algorithm implemented in SimpleElastix \cite{marstal2016simpleelastix, klein2009elastix}, owing to its broad applicability and high accuracy in both MRI and echocardiographic images. It utilizes the Mean Square Error (MSE) as the similarity metric, and transform bending energy as the smoothness regularization. \\

Given 4D images \( I \in \mathbb{R}^{X_i \times Y_i \times Z_i \times T} \), where \( X_i, Y_i, Z_i \) denote the spatial dimensions (\(64 \times 64 \times Z_{\text{slice}}\) in our cases), and \( T \) denotes the number of time points per cardiac cycle (ranging from \( 20 \) to \( 40 \) in our dataset), the resulting B-spline coefficient tensor is denoted by:
\[
C^{\text{coef}} \in \mathbb{R}^{T \times (X_c \times Y_c \times Z_c) \times \text{dim}}.
\]
where \( X_c, Y_c, Z_c \) denote the number of control points along each spatial dimension, and \( {dim} \) indicates the spatial dimensionality of the domain (3 for 3D in this case). The B-spline control points are arranged in a grid of size \( 11 \times 11 \times 11 \), consistent with previous empirical findings~\cite{wiputra2020cardiac}. Registration is conducted between consecutive frames throughout the entire cardiac cycle.\\

The displacements at the next time point \(d\overrightarrow{X}_{t \rightarrow t+\delta t}\) can be calculated by,

\begin{equation}
\displacement = \sum_{l=0}^{3} \sum_{m=0}^{3} \sum_{n=0}^{3} B_{l}(o)B_{m}(p)B_{n}(q)\overrightarrow{C}_{i+l,j+m,k+n}^{\text{coef}}
\label{eq:disp}
\end{equation}

\noindent where ${B}_{n}$ denotes the 3D B-spline basis functions \cite{620490}, and $o$, $p$, and $q$ represent the local coordinates of the B-spline control point grid.

\paragraph{\textbf{Fourier Regularization}:} 
Next, EDFL uses a fully connected neural network (FCNN) for temporal smoothness and cyclic regularization of the pairwise registration. The network takes in image coordinate \( (P, 3) \) (\(P \in \mathbb{R}^{64 \times 64 \times Z_{\text{slice}}}\)) and B-spline coefficients \(C^{\text{coef}}\), outputs Fourier coefficients  \( \overrightarrow{F}_{f} \) (\(\overrightarrow{F}_{f}\in \mathbb{R}^{3} \), with \( f \) specifying the current frequency mode) at each coordinate, expressed as:

\begin{equation}
\overrightarrow{X}_{f}^{t} = \sum_{f=0}^{N} \left( \overrightarrow{F}_{f} \times \cos\left(\frac{2\pi ft}{T}\right) + \overrightarrow{F}_{f+N+1} \times \sin\left(\frac{2\pi ft}{T}\right) \right)
\label{eq:3}
\end{equation}

\noindent where $\overrightarrow{X}_{f}^{t}$ is the displacements over time, $N$ is the number of Fourier terms representing different Fourier frequency (($N=4$ here, in accordance to previous optimization \cite{wiputra2020cardiac}).\\

The loss function is the accumulative difference between Fourier-dictated deformation and Elastix deformation at different time points across time and space,

\begin{equation}
Loss = \frac{1}{P} \sum_{P} \sum_{M} {\left\| \overrightarrow{X}_{f}^{t+\delta t} - \overrightarrow{X}_{f}^{t} - d \overrightarrow{X}_{e}^{t \rightarrow t+\delta t}\right\|_{2}^{2}}
\label{eq:4}
\end{equation}

\noindent where $P$ is the number of coordinates, $M$ represents the number of pairwise groups, $\overrightarrow{X}_{f}^{t+\delta t}$ and $\overrightarrow{X}_{f}^{t}$ are the Fourier displacements at two distinct time points. The term $d \overrightarrow{X}_{e}^{t \rightarrow t+\delta t}$ denotes the pairwise Elastix registration displacements from time point $t$ to $t+\delta t$, calculated by inserting $\overrightarrow{X}_{f}^{t}$ into Eqn.\ref{eq:disp}.

\subsubsection{Datasets and Evaluation}
To evaluate the performance of the proposed EFDL registration framework, we conduct experiments on two datasets: 50 3D+t adult cardiac MRI cases from the ACDC public dataset \cite{8360453}, and 10 3D+t fetal echocardiography cases from a previously published private dataset \cite{wiputra2020cardiac, chan2021full}. No image pre-processing was applied to either dataset. Each case consists of 20 to 40 temporal frames over the cardiac cycle.\\

We compared EFDL with the following registration methods: (1) Elastix, a classical traditional registration approach; (2) Bspline Fourier, an extension of SimpleElastix with temporal Fourier-based regularization for cyclic motion \cite{chan2021full, wiputra2020cardiac}; (3) Voxelmorph, a widely adopted DLIR method for pairwise registration; and (4) Co-attention STN \cite{ahn2023co}, a state-of-the-art (SOTA) DL 4D image registration approach.\\

Here, B-spline Fourier, Co-attention STN, and the proposed EFDL incorporate temporal constraints and are specifically designed for groupwise registration across the cardiac cycle. In contrast, Elastix and VoxelMorph are pairwise registration algorithms that do not enforce temporal consistency across time frames. Additionally, Elastix, B-spline Fourier, and EFDL are not pre-trained, and thus require patient-specific training for each new inference. As a result, they exhibit substantially longer inference times compared to pre-trained DLIR methods such as VoxelMorph and Co-attention STN. However, these approaches do not rely on large-scale datasets of specific image modalities for training, which can be advantageous in data-limited scenarios.\\

During VoxelMorph training, for optimal effects, all possible frame-pair combinations within the cardiac cycle are used (i.e., all-to-all pairs). In contrast, Co-attention STN employs a more selective strategy based on anatomical variability: only frame pairs with a Dice overlap below 0.7 are included in the training (following the approach in \cite{ahn2023co}), to ensure sufficient motion diversity during training. The ED frame serves as the reference to which all other time points are registered. The ratio of training to testing cases is maintained at 4:1 for both adult and fetal datasets.\\

Performance evaluation is carried out by comparing the registered cardiac shape images to ground truth segmentations. For adult datasets, the ACDC myocardium masks are used as ground truths, while for fetal datasets, manual segmentations, performed using LazySnap \cite{wiputra2016fluid}, are used. The quality of alignment is evaluated by Dice Similarity Coefficient (DSC) and Hausdorff Distance (HD) \cite{aydin2021usage}, as defined in supplementary text.

\subsection{Myocardial Stiffness and Maximal Tension Estimator}

The myocardial stiffness and maximum tension estimator are implemented using PINN, which first estimates the coefficients of the stiffness model, followed by the maximum contractile tension. The overall procedure is illustrated in Fig.~\ref{FIG:schematic} II and consists of the following steps.
\begin{enumerate}
    \item \textbf{3D Cardiac Mesh Reconstruction} The 3D surface mesh of the LV myocardium is first reconstructed through manual segmentation of a single reference frame using LazySnap \cite{wiputra2016fluid} for coarse delineation, followed by surface refinement in Geomagic ~\url{https://oqton.com/geomagic-designx}. However, this can similarly be achieved with deep learning segmentation and reconstruction algorithms \cite{ilesanmi2024reviewing,luo2024explicitdifferentiableslicingglobal,van20183d}.The resulting surface mesh is then converted into a tetrahedral FE mesh using Gmsh \cite{geuzaine2009gmsh}. This reference FE mesh is subsequently propagated to all other time points using motion fields estimated by the EFDL framework. In addition, at each time point, the cavity volume is computed by sealing the endocardial surface and performing volumetric integration on the surface mesh. The resulting volume serves as a reference for the next step in performing the volume constraint in the PINN-FE solver.

    \item \textbf{Load-Free State Definition:} The reference zero-pressure unloaded geometry is assumed to correspond to the cardiac state at 1/3 into the diastolic duration, as this phase typically exhibits the lowest blood cavity pressure. This assumption is imperfect, but consistent with prior studies \cite{wenk2012first, gao2014dynamic, di2018gaussian}, and we believe that future work to develop a deep learning tool to predict the true load-free state is possible. At this reference time point, the myocardial helix angle is assumed to vary linearly from the epicardium to the endocardium (values given in supplementary text section S2). The following parameters are precomputed at the load-free state:
    \begin{itemize}
        \item Fiber (\(f_{unloaded}\)), sheet (\(s_{unloaded}\)), and sheet-normal (\(n_{unloaded}\)) orientations. (methods given in supplementary text section S2)
        \item Nodal endocardial area (\(A_{\text{unloaded}}\)) and myocardial volume (\(V_{\text{unloaded}}\)) are computed by distributing the area of each triangular surface element and the volume of each tetrahedral element to their associated nodes via weighted summation \cite{buoso2021personalising}.
        \item Mesh labels (\(Label_{unloaded}\)) [interior: 0, endocardium: 1, epicardium: 2, base: 3] are assigned to each node.
        \item The spatial derivative operator at each node (\(\nabla u\)) is calculated, as explained in Section~2.2.1.
    \end{itemize}

    \item \textbf{Motion Analysis via POD:} A previous PINN FE model \cite{buoso2021personalising} encoded cardiac motion using shape modes derived from Proper Orthogonal Decomposition (POD) applied to cardiac geometries reconstructed from multiple subjects at a fixed phase of the cardiac cycle. However, this approach captures inter-subject anatomical variability rather than intra-subject dynamic deformations. Here, we apply POD to the series of cardiac shapes across the cardia cycle from a single subject. This enables the extraction of subject-specific dominant motion modes, which are then used for compact and physiologically meaningful motion encoding. The resulting motion modes are used as a reduced-order representation of displacement fields in the PINN-FE solver, which improves numerical stability and accelerates convergence. \\
    
    \item \textbf{Stiffness and Maximal Tension Estimation:} Two similar neural networks are trained to estimate myocardial stiffness (\(C_{\text{stiffness}}\)) and maximum active tension (\(T_{\text{max}}\)), using motion mode amplitudes at ED and PS as inputs. The physical constraints incorporated into the loss functions are presented in Section~2.2.3, and the two networks are described in detail in Section~2.2.4.
\end{enumerate}
\subsubsection{The Spatial Derivative Operator \(\nabla u\) }
The spatial derivative operator \(\nabla u\) acts as a fundamental differential operator that maps the network-predicted displacement field to deformation gradients, and is employed in the calculation of physics constraints.\\

This operator is computed at each node based on the method described by Mancinella et al. \cite{mancinelli2019comparison}, which estimates the gradient of a scalar function $f(x)$ defined over tetrahedral meshes, representing displacement field vector in our case. For each element, the gradient is first evaluated at its centroid $\nabla f(\mathbf{x}_c) = \left( \nabla f(\mathbf{x}_c)_x, \nabla f(\mathbf{x}_c)_y, \nabla f(\mathbf{x}_c)_z \right)$ by solving a local linear system constructed from the coordinates and function values of the element’s four vertices $v_j,v_i,v_k,v_l$ as,
\begin{equation}
    \begin{bmatrix}
        v_j - v_i \\
        v_k - v_i \\
        v_l - v_i
    \end{bmatrix}
    \begin{bmatrix}
        \nabla f(\mathbf{x}_c)_x \\
        \nabla f(\mathbf{x}_c)_y \\
        \nabla f(\mathbf{x}_c)_z
    \end{bmatrix}
    =
    \begin{bmatrix}
        f(x_j) - f(x_i) \\
        f(x_k) - f(x_i) \\
        f(x_l) - f(x_i)
    \end{bmatrix},
\end{equation}

The nodal gradient \(\nabla u\) is then obtained as the volume-weighted average of the gradients from all neighboring tetrahedral elements that share the node. These local derivative matrices can be precomputed and assembled into a global matrix operator outside the network, allowing efficient evaluation of nodal gradients via matrix–vector multiplication when using the network. To avoid numerical instability, elements involving apex nodes are excluded from the assembly due to their small size and disproportionate influence \cite{buoso2021personalising}.

\subsubsection{Proper Orthogonal Decomposition (POD)}
The use of POD shape modes in IMC-PINN-FE is inspired by previous work \cite{buoso2019reduced,manzoni2012model,quarteroni2007numerical,rowley2005model}. Before applying POD, all meshes undergo spatial alignment for consistency.Specifically, each mesh is translated such that the center of the basal plane aligns with the origin of the Cartesian coordinate system. The mesh is then rotated to orient the basal plane to be normal to the positive \(z\)-axis.\\

The mesh size for an individual's LV is selected to contain between 2,000 and 3,000 nodes, balancing computational efficiency with low discretization error. A total of 20 to 40 temporal frames (depending on data acquisition protocols) across the cardiac cycle are included in the POD computation. Displacement is defined as the deviation of each node from its position in the identified load-free state. These displacement fields are concatenated across time into a high-dimensional data matrix, and POD is applied via singular value decomposition (SVD) to extract a set of orthonormal spatial basis functions \cite{buoso2019reduced,buoso2021personalising}. The resulting basis functions are ranked in descending order of their contribution to the total variance, and the first 10 dominant modes are retained to represent the subject-specific deformation subspace. The displacement at any time point, denoted as \(\mathbf{u}\), is approximated as a linear combination of these basis functions: 
\begin{equation}
u = \Psi \overrightarrow{A}_{n,t}
\label{equ:mode}
\end{equation}

\noindent where \(\Psi = [\Psi_1, \dots, \Psi_n]\) represents the set of POD basis functions (in this project, we use the first 20 bases), and the term \(\overrightarrow{A}_{n,t}\) represents the amplitudes corresponding to each basis function.  \\

\subsubsection{Physical Constraints in LV Loss Function}
The physics loss (governing equation) is derived from the potential energy function that describes mechanical behavior across the cardiac cycle, as proposed by \cite{shavik2018high}:
\begin{equation}
loss_{energy} = \int_{\Omega_{\text{myo}}} W \, dV - \int_{\Omega_{\text{endo}}} P \cdot u \, dA 
\label{equ:energy loss}
\end{equation}

\noindent The first term represents the overall strain energy of the myocardium, where \(V\) is the myocardial volume. The second term represents the work done by the endocardium on blood via pressure, \(P\) is the pressure in the LV cavity, and \(u\) is the endocardial displacement relative to the unloaded (reference) state, while \(A\) is the endocardial surface area.\\ 

$W$ is the strain energy density, calculated by:
\begin{equation}
W = W_{passive}+W_{active}
\label{equ:strain energy}
\end{equation}

\noindent where $W_{passive}$ and $W_{active}$ refer to passive and active components, respectively. $W_{passive}$  is calculated using a strain energy function for a Fung-type transversely-isotropic hyperelastic material \cite{guccione1991passive}:
\begin{equation}
W_{passive} = \frac{1}{2} C_\text{stiffness} \left( e^Q - 1 \right)
\end{equation}
where,
\begin{align}
Q &= b_f E_{ff}^2 + b_{xx} \left( E_{ss}^2 + E_{nn}^2 + E_{sn}^2 + E_{ns}^2 \right) \notag \\
&\quad + b_{fx} \left( E_{fn}^2 + E_{nf}^2 + E_{fs}^2 + E_{sf}^2 \right)
\end{align}
% \begin{equation}
% E=\frac{1}{2}(F^TF-I)
% \end{equation}

\noindent where \( E_{ij} \) \( (E=\frac{1}{2}(F^TF-I)) \) represents the Green-Lagrange strain tensor, with subscripts \( f, s, \) and \( n \) indicating the myocardial fiber, sheet, and sheet-normal directions, respectively. \( b_{ij} \) are the stiffness parameters. \( C_{\text{stiffness}} \) is the myocardial stiffness, as estimated by the stiffness estimator. \( F \) is the deformation gradient tensor, defined as $F = \frac{\partial x}{\partial X} = I + \nabla u  \cdot u $, where \( x \) represents the deformed coordinates, and \( X \) corresponds to the unloaded (reference) coordinates. \\

According to \cite{moose_strain_energy_density}, the active energy density $W_{active}$ can be expressed as:
\begin{equation}
 W_{\text{active}} = \int_0^E \sigma_{\text{active}} : dE 
 \end{equation}

 \noindent where $\sigma_{\text{active}}$ is the Cauchy stress tensor, which is determined using a previously established calcium activation model \cite{guccione1991passive}, characterizing the sigmoidal relationship between chemical activation and cardiac muscle tension:
 \begin{equation}
\sigma_{\text{active}} = T_{\max} \frac{Ca_0^2}{Ca_0^2 + ECa_{50}^2} C_t 
 \end{equation}
 
\noindent where \( T_{\max} \) represents the maximum active tension, as estimated by the maximal tension estimator. The terms \( Ca_0 \), \( ECa_{50} \), and \( C_t \) characterize the calcium activation dynamics, with \( Ca_0 \) denoting the peak calcium concentration, \( ECa_{50} \) indicating the sarcomere length-dependent calcium sensitivity, and \( C_t \) capturing temporal variations:
\begin{equation}
ECa_{50} = \frac{(Ca_0)_{\max}}{\sqrt{e^{B (l - l_0)} - 1}}, \quad
C_t = \frac{1}{2} (1 - \cos \omega)
\end{equation}

\noindent where \(B\) is a constant, \((Ca_0)_{\max}\) is the maximum peak intracellular calcium concentration, and \(l_0\) is the sarcomere length at which no active tension develops. \( \omega \) depends on the cycle time and varies as follows:

\begin{equation}
\omega =
\begin{cases} 
    \pi \frac{t}{t_0} & \text{when } 0 \leq t < t_0, \\
    \pi \frac{t - t_0 + t_r}{t_r} & \text{when } t_0 \leq t < t_0 + t_r, \\
    0 & \text{when } t_0 + t_r \leq t.
\end{cases}
\end{equation}

\noindent where \( t_0 \) represents the time to peak tension, indicating the duration required for myofiber contraction, while \( t_r \) denotes the relaxation time \cite{guccione1991passive}, defined as:
\begin{equation}
t_r=ml+b
\end{equation}

\noindent where \( m \) represents the slope of the linear relaxation duration-sarcomere length relation, \( b \) is the time-intercept of this relation, and \( l \) is the sarcomere length, which varies with myofiber stretch \(\lambda\). 
\begin{equation}
\lambda = \sqrt{e_{f0} \cdot C e_{f0}},
\quad l = \lambda l_r
\end{equation}
where \(l_r\) is the relaxed sarcomere length, \(e_{f0}\) defines the fiber direction in the reference configuration.\\

The overall Cauchy stress \cite{avril2023inverse} can be calculated by:
\begin{equation}
\sigma = \frac{2}{J} F \frac{\partial W}{\partial \left( F^T F \right)} F^T
\label{stress}
\end{equation}
where $J = \det(F)$ is the determinant of the deformation gradient $F$.\\

All predefined parameters in these equations, based on the literature, are shown in supplementary S2.\\

\subsubsection{Estimation Framework of \(C_\text{stiffness}\) and \(T_\text{max}\)}

\paragraph{\textbf{Stiffness Estimation}:} 

The stiffness estimator network takes as input the amplitudes of the POD modes extracted from the ED frame, and outputs the myocardial stiffness coefficient \( C_{\text{stiffness}} \), while the POD motion bases and the LV cavity pressure at ED are treated as predefined parameters. The stiffness coefficient represents the global stiffness scaling factor, and the optimization is conceptually akin to inferring this factor by simulating the diastolic pressure loading process that deforms the ventricle from the load-free state to the ED configuration. For this estimation task, only the passive component of the strain energy is considered (\( W = W_{\text{passive}} \)), active contraction is assumed to be absent.\\

Clinically, direct measurement of LV ED pressure (LVEDP \(P_{ED}\)) is challenging, as it typically requires invasive catheter-based procedures. However, empirical equations derived from prior clinical studies enable estimation of \(P_{ED}\) using non-invasive imaging data. For cardiac MRI, we apply empirical formulations provided by Burkhoff et al.\cite{burkhoff2005assessment}:
\[
P_{ED} = 0.015  \left(e^{0.06 (V_{ED} - V_0)} - 1 \right)
\]

\noindent where \( V_{\text{ED}} \) is the ED volume obtained from MRI, and \( V_0 \) is the theoretical unloaded volume of the LV at zero pressure.\\

For echocardiography, \(P_{\text{ED}}\) is estimated based on the ratio between mitral inflow Doppler velocity (\(E\)) and mitral annular early diastolic Doppler velocity (\(E'\)), as proposed by \cite{zheng2022morphological}:

\begin{equation}
P_{ED} = 0.85 \times \frac{E'}{E} + 4.4
\end{equation}

The computed stiffness coefficient is used for maximum tension estimation in the next network.

\paragraph{\textbf{Maximal Tension Estimation}:} 
The maximal tension estimator takes inputs of the amplitudes of the POD modes extracted from the PS frame and outputs the maximal active tension \(T_{max}\), while the POD motion basis, stiffness coefficient, and LV cavity pressure at PS are treated as predefined parameters. Here, we assume that the timing of the PS frame is known, practically, it can be obtained from tonometry or photoplethysmography combined with ECG. At this stage, both active and passive components of the strain energy are considered, and the total strain energy is computed according to Eq.~\ref{equ:strain energy}. Practically, the LV pressure at PS can non-invasively measured from the upper arm via a standard pressure cuff. \\

\subsubsection{Datasets and Evaluation}
\begin{figure}[htbp]
	\centering
	\includegraphics[width=0.7\textwidth]{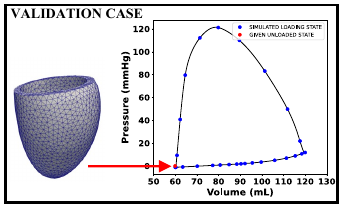}
    \caption{Schematic of validation case generation: A limited number of states (blue points) are selected from the FE simulation, which is performed using a given unloaded surface mesh (red points), to serve as tracking results.}
	\label{FIG:val}
\end{figure}

To evaluate the accuracy of the back-computed parameters ($C_{\text{stiffness}}$ and $T_{\max}$), we performed FE simulations using predefined values of $C_{\text{stiffness}}$ and $T_{\max}$, which served as ground truths. From these simulations, a limited number of time frames are extracted and used as cardiac motion estimation inputs in the estimators (as illustrated in Fig.~\ref{FIG:val}). The estimated $C_{\text{stiffness}}$ and $T_{\max}$ values were then compared against the ground truth to assess the accuracy of the estimator.

In total, 8 LV FE simulations were performed. 2 were based on healthy adult human image data from the \textit{Cardiac Atlas Project}~\url{https://www.cardiacatlas.org/}, 2 were based on healthy fetal echocardiography images from our private dataset~\cite{wiputra2020cardiac, chan2021full}, and 4 were based on porcine echocardiography images from a previous study~\cite{zheng2022morphological}, including two healthy pig hearts and two with LV hypertrophy induced by surgical aortic constriction. These simulations therefore span a range of LV sizes and include both healthy and pathological conditions.\\

In addition to the FE-based validation, we further tested our estimator on 2 cases of real clinical images, using EFDL for motion tracking: 1 adult case MRI from the \textit{Cardiac Atlas Project} and 1 fetal case from our private dataset \cite{wiputra2020cardiac}. Since the ground truth myocardial stiffness and active tension are not known, we adopted values from literature \cite{arts1979model, shavik2018high, green2023dependency} as ground truths. These are not ideal ground truths, but applying the framework to these real cases serves to demonstrate its practical feasibility.

\subsection{Volume-Constrained PINN-FE Solver}

\begin{figure*}[htbp]
	\centering
	\includegraphics[width=0.98\textwidth]{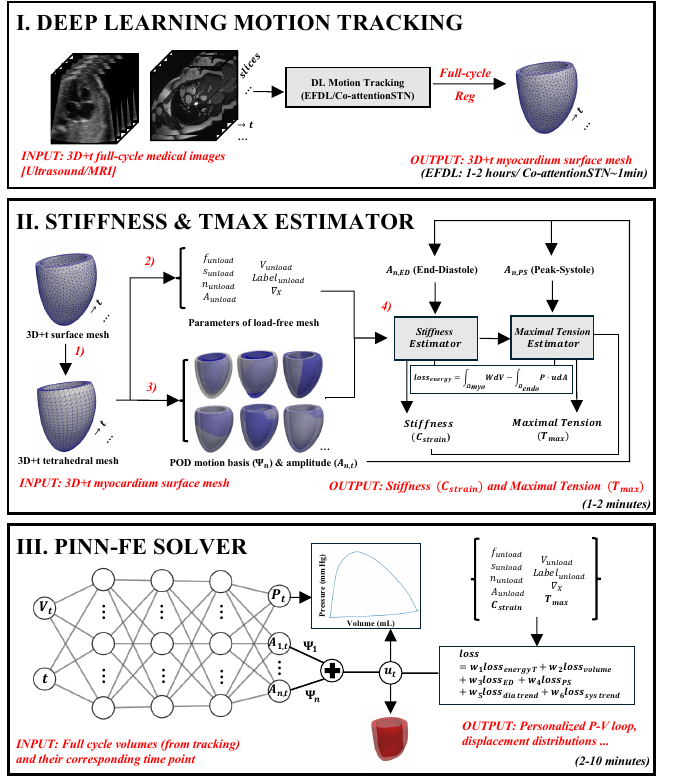}
    \caption{
    Workflow of the IMC-PINN-FE framework. 
    \textbf{I. Deep Learning Motion Tracking:} 4D full-cycle cardiac images (Echo/MRI) are used to generate myocardial motion meshes via either EFDL (our proposed method, no pre-training required) or Co-attention STN (previous SOTA method, pre-training required). 
    \textbf{II. Myocardial Stiffness and Maximal Tension Estimator:} Starting from the 3D+t surface mesh, the pipeline includes: (1) tetrahedral meshing via Gmsh; (2) preprocessing to define helix angles and the load-free reference state; (3) POD to extract motion bases and amplitudes across frames; and (4) personalized estimation of myocardial stiffness (\(C_{\text{stiffness}}\)) and maximal active tension (\(T_{\text{max}}\)) using ED and PS amplitudes. 
    \textbf{III. Volume-Constrained PINN-FE Solver:} Inputs include myocardial volumes and their time points (from Section~2.1), sampled uniformly from the cardiac cycle starting at ED. A PINN with MLP architecture predicts LV pressure \( P_t \) and motion amplitudes \( A_{n,t} \), from which displacement fields \( u_t \) are reconstructed via Eq.~\ref{equ:mode}, enabling personalized P–V loop and biomechanical analysis.
    }
	\label{FIG:schematic}
\end{figure*}

The personalized PINN-FE solver is employed to model LV biomechanics over the entire cardiac cycle (Fig~\ref{FIG:schematic} III), utilizing the encoded motion modes and the personalized stiffness and active tension parameters estimated in the previous step. The solver incorporates a FCNN to predict the LV pressure \( P_t \) and mesh displacements \( u_t \) at each time point \( t \) within the cardiac cycle, using the elapsed time since ED (\( t \)) and the corresponding LV cavity volume \( V_t \) as inputs. Both \( t \) and \( V_t \) are uniformly sampled across the cardiac duration, based on coarse frames derived from the tracking results in Section~2.1. In our implementation, the cardiac duration is assumed to be 800~ms for adults \cite{shavik2018high} and 400~ms for fetuses \cite{green2023dependency}. These durations can be adapted in future applications to accommodate patient-specific timing. The node displacement outputs are in the form of amplitudes of the motion basis \( A_{n,t} \), and can be decoded as per Eq.~\ref{equ:mode}. Subsequently, stress can also be computed via Eq.~\ref{stress}. IMC-PINN-FE is volume-constrained in the sense that it directly adopts the LV volume over time obtained from the images, and only computes the corresponding pressure. It does not interface with a lumped parameter model to determine the LV volume. However, such a model can easily be incorporated in future.

\subsubsection{Loss function for PINN-FE solver}

The loss function of this network consists of six components: 
\begin{enumerate}
    \item The physics-governing equation, Eq.~\ref{equ:energy loss2}.
    \item The volume control equation, which ensures that LV volume is consistent with images, Eq.~\ref{equ:volume}.
    \item The ED pressure control equation, which enforces the ED pressure at the ED frame, Eq.~\ref{equ:ED/PS}.
    \item The PS pressure control equation, which enforces the PS pressure at the PS frame, Eq.~\ref{equ:ED/PS}.
    \item The diastolic phase slope constraint, which ensures a continuous increase in the P-V loop slope during diastole to aid convergence, Eq.~\ref{equ:slope}.
    \item The systolic phase slope constraint, which ensures a continuous increase in the P-V loop slope to aid convergence, Eq.~\ref{equ:slope}.
\end{enumerate}

\begin{equation}
loss_{energy \ T} = \frac{1}{T}\sum_T(\int_{\Omega_{\text{myo}}} W \, dV - \int_{\Omega_{\text{endo}}} P \cdot u \, dA)^2
\label{equ:energy loss2}
\end{equation}

\begin{equation}
\mathit{loss}_{\mathit{volume}} = \frac{1}{T} \sum_{t} {(V_{\mathit{predict}} - V_{\mathit{target}})^2}
\label{equ:volume}
\end{equation}

\begin{equation}
\mathit{loss}_{\mathit{ED/PS}} = {(P_{\mathit{ED/PS, predict}} - P_{\mathit{ED/PS, target}})^2}
\label{equ:ED/PS}
\end{equation}

\begin{equation}
\mathit{loss}_{\mathit{dia/sys \ trend}} = \sum_{T_{\mathit{dia/sys}}} \max\bigg( 0,\frac{\Delta P_{t_{\mathit{dia/sys}}}}{\Delta V_{t_{\mathit{dia/sys}}}} -\frac{\Delta P_{t_{\mathit{dia/sys}}+1}}{\Delta V_{t_{\mathit{dia/sys}}+1}} \bigg)
\label{equ:slope}
\end{equation}
where \( T \) represents the total number of time points,  \( V_{\mathit{predict}} \) and \( V_{\mathit{target}} \) denote the predicted and target myocardial volumes, respectively,  \( P_{\mathit{ED/PS, predict}} \) and \( P_{\mathit{ED/PS, target}} \) represent the predicted and target LV pressures at the ED/PS frame, and \( \frac{\Delta P_{t_{\mathit{dia/sys}}}}{\Delta V_{t_{\mathit{dia/sys}}}} \) denotes the slope of P-V loop in the diastolic or systolic phase.\\

Hence, the overall loss function is:
\begin{align}
\mathit{loss} &= w_1 \mathit{loss}_{\mathit{energy \ T}} + 
w_2 \mathit{loss}_{\mathit{volume}} +\notag \\
&\quad w_3 \mathit{loss}_{\mathit{ED}} + 
w_4 \mathit{loss}_{\mathit{PS}} + \notag \\
&\quad w_5 \mathit{loss}_{\mathit{dia \ trend}} + 
w_6 \mathit{loss}_{\mathit{sys \ trend}}
\label{equ:LOSS_all}
\end{align}
where \( w_i \) represents the weight of each component, as defined in supplementary S3.\\

The comparison group is the set of volume-constrained LV FE simulations, generated based on our previous work using the same equations as in Section~2.2.3. This traditional FE source is available at \url{https://github.com/WeiXuanChan/heartFEM}. 

\clearpage

\subsubsection{Datasets and Evaluation}
To evaluate the performance of volume-constrained PINN-FE solver, we conducted the complete procedure of IMC-PINN-FE in Fig.~\ref{FIG:schematic} to model cardiac biomechanics across the entire cardiac cycle. We compared IMC-PINN-FE with the traditional FE simulation and the previously proposed PINN-FE method by Buoso et al.~\cite{buoso2021personalising}. We assess if the simulations can maintain a closeness to the imaged motions, and if they can produce realistic pressure-volume (P-V) loops.\\

This evaluation was conducted on 6 human cases introduced in Section~3.2, comprising 3 adult cardiac MRI cases and 3 fetal echocardiography cases. The ground truth for imaged cardiac motion was defined as 4D LV mesh reconstructions propagated across the cardiac cycle using EDFL due to its highest accuracy (see Table~\ref{tab:registration_results}) without requiring a large training dataset.

\subsection{Neural Network Architecture}
For section 2.1, all EFDL experiments were conducted using the NVIDIA Modulus framework \cite{NVIDIAModulus} (PyTorch v1.4.0). The model was trained using a multilayer perceptron (MLP) architecture consisting of 4 layers, each containing 1024 neurons. All EFDL models were optimized using the Adam optimizer with a learning rate of 0.001 and a decay rate of 0.95.\\

For section 2.2, both estimators employ a MLP with 3 hidden layers, each containing 32 neurons, and are trained to converge by minimizing Eq.~\ref{equ:energy loss} to zero. Both estimators were optimized using the AdamW optimizer with a learning rate of 0.001. \\

For section 2.3, the PINN-FE neural network is configured with 4 hidden layers, each containing 128 neurons, and is trained to converge by minimizing Eq.~\ref{equ:LOSS_all}. The framework is optimized using the AdamW optimizer with a learning rate of 0.0001 and a decay rate of 0.8. \\

All models were executed on NVIDIA GeForce RTX 3090 Ti PCIe 3.0 GPUs with 24GB GDDR6X memory.

%% file: Results.tex
\subsection{Performance of EFDL}

Table~\ref{tab:registration_results} presents the quantitative results of different registration methods on the Adult MRI and Fetal Echo datasets. The evaluation metrics, DSC and HD, are first averaged over the entire cardiac cycle for each case, and then reported as the mean and standard deviation across all cases.\\

For the Adult MRI dataset, our EFDL achieves the highest DSC ($0.829 \pm 0.026$) and the lowest HD ($4.71 \pm 0.17$), demonstrating superior registration accuracy. Co-attention STN and B-spline Fourier yield comparable performance (with DSC differences within 0.014 and HD within 0.19). However, B-spline Fourier requires at least 2.4 times longer training time. As for Co-attention STN, although it can be used for inference within a very short time after training, it must be retrained when applied to different image type or modality. Such training is time-consuming (Table~\ref{tab:registration_results}), and requires a large dataset with ground truths, which can be cumbersome to prepare. Further,  image down-sampling is required due to GPU memory constraints, which in turn limits the achievable registration accuracy. On the other hand, EFDL does not require lengthy dataset preparation and pre-training, but it requires training for every new subject, which is typically within 1.6 hours per case.\\

For the Fetal Echo dataset, Co-attention STN achieves the highest DSC and lowest HD. EFDL demonstrates the next best performance, with comparable results, achieving a DSC of $0.802 \pm 0.030$ and an HD of $4.99 \pm 0.45$.\\

Overall, both Co-attention STN and EDFL demonstrate strong performance, each suited to different application needs. Co-attention STN, which benefits from extensive pretraining and is tailored for large-scale, single-modality images, is a suitable choice when sufficient training data and computational resources are available, as it can give real-time results. However, in scenarios of limited data availability, EDFL offers a more practical solution. It has a good balance of accuracy and computational efficiency, and can be a good choice in such scenarios. All subsequent experiments in this paper are conducted using EDFL as the registration method.

\begin{table}[h]
    \centering
    \footnotesize
    \caption{Quantitative comparisons of various 3D cardiac motion estimation techniques on the AdultMRI and FetalEcho datasets. Methods marked with * incorporate temporal consistency constraints, while methods marked with $\dagger$ do not need pre-training with a large dataset. t denotes the average inference time. The best results are highlighted in \textbf{\textcolor{red}{red}}.}
    {\rmfamily % Apply Times New Roman only to the table
    \begin{tabular}{ l c c c c }
        \rowcolor{gray!30} 
        \textbf{Method} & \textbf{Dataset} & \textbf{DSC $\uparrow$} & \textbf{HD(mm) $\downarrow$} & \textbf{t(h) $\downarrow$} \\ 
        \hline
        No registration & MRI & 0.681±0.109 & 6.61±1.12 & - \\ 
        & Echo & 0.623±0.120 & 6.83±1.61 & - \\ 
        \hline
        $\dagger$Elastix  & MRI & 0.736±0.112 & 5.97±1.33 & \textbf{\textcolor{red}{0.4}} \\ 
        \cite{marstal2016simpleelastix} & Echo & 0.729±0.108 & 6.01±1.05 & \textbf{\textcolor{red}{0.4}} \\ 
        \hline
        $\dagger$*Bspline Fourier & MRI & 0.815±0.030 & 4.86±0.34 & 3.9 \\ 
        \cite{chan2021full, wiputra2020cardiac} & Echo & 0.803±0.008 & 4.94±0.26 & 4.0 \\ 
        \hline
        VoxelMorph & MRI & 0.766±0.067 & 5.85±1.10 & 8.1 \\ 
        \cite{balakrishnan2019voxelmorph} & Echo & 0.703±0.051 & 6.11±0.96 & 8.3 \\ 
        \hline
        *Co-attentionSTN & MRI & 0.817±0.031 & 4.90±0.22 & 7.4 \\ 
        \cite{ahn2023co} & Echo & \textbf{\textcolor{red}{0.804±0.012}} & \textbf{\textcolor{red}{4.92±0.19}} & 7.2 \\ 
        \hline
        $\dagger$*EFDL & MRI & \textbf{\textcolor{red}{0.829±0.026}} & \textbf{\textcolor{red}{4.71±0.17}} & \textbf{1.6} \\ 
        & Echo & 0.802±0.030 & 4.99±0.45 & \textbf{1.7} \\ 
        \hline
    \end{tabular}
    }
    \label{tab:registration_results}
\end{table}

\subsection{Performance of $C_\text{stiffness}$ and $T_\text{max}$ Estimator}

\begin{table*}[htbp]
\caption{Testing results of the \(C_{\text{stiffness}}\) and \(T_{\max}\) estimators across 2 applications on real clinical images and 8 applications on FE-simulated validation cases. EDV and ESV are derived from 3D reconstructions from images. For validation cases, EDP, PSP, reference \(C\), and reference \(T_{\max}\) are reasonable values assumed in the FE simulations, based on the literature~\cite{ong2021biomechanics,wang2010cardiac,muller2018effects,gao2013electro,shavik2018high,racca2016contractile}. For image tracking cases, they are based on literature values \cite{green2023dependency,racca2016contractile,wang2010cardiac,muller2018effects,shavik2018high,zheng2022morphological}. RE denotes the relative error between the predicted values (PINN) and the reference values (literature range or FE settings). The prediction time of the network is 63 ± 22 seconds per case.}
\begin{adjustbox}{max width=\textwidth}
{\rmfamily
\begin{tabular}{cccccccc
>{\columncolor[HTML]{E0E0E0}}c cc
>{\columncolor[HTML]{E0E0E0}}c }
\rowcolor{gray!40} 
\textbf{Source}                                                                       & \textbf{Case} & \textbf{\begin{tabular}[c]{@{}c@{}}EDV \\(mL)\end{tabular}} & \textbf{\begin{tabular}[c]{@{}c@{}}ESV \\(mL)\end{tabular}} & \textbf{\begin{tabular}[c]{@{}c@{}}EDP \\(mmHg)\end{tabular}} & \textbf{{\begin{tabular}[c]{@{}c@{}}PSP \\(mmHg)\end{tabular}}} & \textbf{\begin{tabular}[c]{@{}c@{}}C (kPa) \\ (PINN)\end{tabular}} & \textbf{\begin{tabular}[c]{@{}c@{}}C (kPa)\\ (LITER/FE)\end{tabular}} & \textbf{RE (\%)$\downarrow$} & \textbf{\begin{tabular}[c]{@{}c@{}}Tmax (kPa)\\ (PINN)\end{tabular}} & \textbf{\begin{tabular}[c]{@{}c@{}}Tmax (kPa)\\ (LITER/FE)\end{tabular}} & \multicolumn{1}{l}{\textbf{RE (\%)$\downarrow$}} \\ \hline
                                                                                      & Adult-1       & 120.88          & 71.31          & 11.0 \cite{fowler1980cardiac,msd2025pressures}
                                                                                      & 120.0 \cite{fowler1980cardiac,msd2025pressures}           & 0.095                                                              & 0.1 \cite{shavik2018high}                                                          & 5.0            & 124.8                                                              & 95-200.7 \cite{wang2010cardiac,muller2018effects,gao2013electro,shavik2018high}                                                         & within                                                       \\ \cline{2-12} 
\multirow{-2}{*}{\begin{tabular}[c]{@{}c@{}}\textbf{Image tracking}\\ (Real)\end{tabular}}     & Fetal-1       & 0.41             & 0.22             & 5.0  \cite{johnson2000intracardiac}             & 27.0 \cite{johnson2000intracardiac}              & 0.094                                                              & 0.1 {}\cite{ong2021biomechanics}{}                                                           & 6.0          & 36.9                                                               & 23.9-59.2 {\cite{racca2016contractile}}                                                        & within                                                       \\ \hline
                                                                                      & Adult-2       & 121.12           & 60.01            & 11.0            & 120.0       & 0.098                                                              & 0.1                                                                   & 2.0            & 103.0                                                             & 100.3                                                               & 2.73                                                       \\ \cline{2-12} 
                                                                                      & Adult-3       & 135.36           & 54.23           & 11.0            & 120.0         & 0.095                                                              & 0.1                                                                & 5.0            & 100.3                                                              & 100.3                                                                  & 0.05                                                       \\ \cline{2-12} 
                                                                                      & Fetal-2       & 0.39             & 0.20             & 2.0              & 21.8            & 0.098                                                              & 0.1                                                                 & 2.0            & 28.8                                                           & 30.0                                                                 & 4.11                                                       \\ \cline{2-12} 
                                                                                      & Fetal-3       & 0.63             & 0.28             & 2.0              & 32.3         & 0.101                                                              & 0.1                                                                & 1.0            & 38.8                                                             & 40.0                                                                 & 3.02                                                       \\ \cline{2-12} 
                                                                                      & PIG-healthy1  & 30.56             & 15.18             & 12.6               & 85.2             & 0.089                                                              & 0.09                                                                   & 1.1            & 42.7                                                             & 40.5                                                                   & 5.41                                                       \\ \cline{2-12} 
                                                                                      & PIG-healthy2  & 34.42            & 15.29            & 10.6               & 86.2               & 0.093                                                              & 0.09                                                                   & 3.3           & 38.6                                                               & 40.5                                                                 & 4.62                                                       \\ \cline{2-12} 
                                                                                      & PIG-LVH1      & 40.01            & 26.51          & 14.3         & 165.2           & 0.377                                                              & 0.4                                                             & 5.8            & 60.2                                                             & 60.0                                                                & 0.40                                                       \\ \cline{2-12} 
\multirow{-8}{*}{\begin{tabular}[c]{@{}c@{}}\textbf{FE tracking} \\ (Validation)\end{tabular}} & PIG-LVH2      & 55.23            & 35.34            & 13.9              & 178.2             & 0.409                                                              & 0.4                                                                 & 2.3             & 58.6                                                               & 60.0                                                               & 2.26                                                       \\ \hline
\end{tabular}
}
\end{adjustbox}
\label{tab:testing_results}
\end{table*}

Table~\ref{tab:testing_results} presents the performance of the \(C_{\text{stiffness}}\) and \(T_{\max}\) estimators on 2 image tracking cases and 8 FE simulation tracking cases. For the FE cases, the ground truth reference values for \(C_{\text{stiffness}}\) and \(T_{\max}\) are those used in the FE simulation settings, which are based on ~\cite{green2023dependency,racca2016contractile,wang2010cardiac,muller2018effects,shavik2018high,zheng2022morphological}. For the image tracking cases, no ground truth is available, but the reference values are taken from the values or ranges of values reported in the literature~\cite{ong2021biomechanics,wang2010cardiac,muller2018effects,gao2013electro,shavik2018high,racca2016contractile}. For \(C_{\text{stiffness}}\), we used values from previous FE simulations that successfully replicated in vivo cardiac function and that used the same stiffness model as us. For adult \(T_{\max}\), a range of literature is used, including those performing inference from medical images and simulations that successfully replicated the cardiac function. For fetal \(T_{\max}\), literature on experimental measurements is used.\\

For the image tracking cases, the estimated \(C_{\text{stiffness}}\) and \(T_{\max}\) demonstrate reasonable agreement with these literature values, or falls within the literature range. For the FE tracking cases, the estimated \(C_{\text{stiffness}}\) and \(T_{\max}\) show strong agreement with the reference values, exhibiting relatively low relative error (RE) across different subject groups. The RE for both \(C_{\text{stiffness}}\) and \(T_{\max}\) remains below 6.0\% in all adult, fetal, and pig FE cases, indicating reasonable predictive performance across a range of physiological conditions. \\

Compared to the traditional approach of performing iterative FE simulations to back-compute \(C_{\text{stiffness}}\) and \(T_{\max}\), our PINN estimators achieve a tremendous reduction in computational time. A typical iterative FE simulation requires several hours for a single iteration (and one case usually requires multiple iterations for converging), whereas our network completes the estimation in 63~±~22 seconds per case.\\

Overall, the results demonstrate the reasonable accuracy of our \(C_{\text{stiffness}}\) and \(T_{\max}\) estimators, along with a significant reduction in computational cost compared to the traditional iterative FE approach, highlighting the potential of IMC-PINN-FE as a real-time and clinically adoptable solution for biomechanical parameter estimation.

\subsection{Performance of PINN-FE solver}

\subsubsection{Results of P-V loop}

In this section, we compare the P–V loops obtained from whole cardiac cycle simulations, using the PINN-FE solver (III of IMC-PINN-FE), the traditional volume-constrained FE model, a state-of-the-art past LV PINN-FE implementation by Buoso et. al.~\cite{buoso2021personalising}, and a fixed image tracking approach. Two representative image cases are used for this (Adult-2 and Fetal-2).\\

IMC-PINN-FE and volume-constrained FE simulations are conducted as described in the methods section. The fixed image tracking method is essentially IMC-PINN-FE but where input displacement fields are fixed as obtained from the images, and are not subject to optimization. A neural network is trained solely to fit pressure values at each time point, using the fixed displacements (represented by motion mode amplitudes) as input. The loss function follows the same formulation as IMC-PINN-FE (Eq.~\ref{equ:LOSS_all}), except that the volume constraint term ($\text{loss}_{\text{volume}}$) is excluded, since volume is already determined by the fixed displacements. This design isolates the pressure fitting process, allowing us to assess whether the unmodified image-derived motion fields, assumed to be error free, can satisfy the physical energy balance, and evaluates whether imaged displacements are consistent with biophysics. 
\\

Buoso et. al.’s PINN-FE adopts a PINN framework that predicts myocardial motion from pressure and active tension inputs. Unlike our method, which uses patient-specific motion fields to encode shape modes, Buoso et. al. uses multi-subject dataset and inter-patient shape variability to encode shape modes, and we use their published modes for implementing their approach. For inference, Buoso et. al. uses pressure and active tension waveforms derived from a Windkessel model to generate myocardial deformation over the cardiac cycle. At present, we have not implemented a Windkessel model for IMC-PINN-FE, but relies on imaged LV volumes rather than Windkessel computations as inputs. To enable comparisons between the two methods, we amended Buoso et. al.’s code to exclude the Windkessel model, using the FE simulation pressures instead of pressure from the Windkessel model as inputs.\\

\begin{figure*}[htbp]
	\centering
	\includegraphics[width=\textwidth]{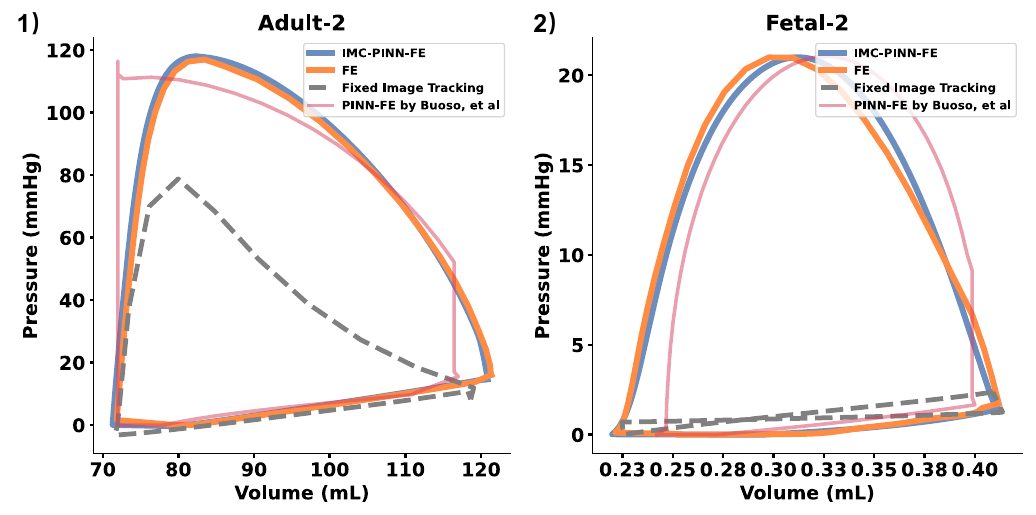}
    \caption{P–V loop evaluation for (1) adult-2 and (2) fetal-2 cases, comparing IMC-PINN-FE, FE simulation, fixed image tracking, and Buoso et al.'s PINN-FE~\cite{buoso2021personalising}. Blue represents the P–V loop predicted by IMC-PINN-FE, orange represents the P–V loop from FE simulation, gray represents the P–V loop generated by fixed displacements from image tracking, and red represents the P–V loop predicted by Buoso et al.'s PINN-FE \cite{buoso2021personalising}.}
	\label{FIG:PV}
\end{figure*}

The results are shown in Fig.~\ref{FIG:PV}. In both image cases, IMC-PINN-FE achieves P–V loops (blue) that are nearly identical to those obtained from the traditional volume-constrained FE simulations (orange), suggesting that IMC-PINN-FE effectively integrated biomechanics principles despite maintaining effective data-driven constraints. Meanwhile, IMC-PINN-FE also enables better match with imaged motion fields, as will be demonstrated in Section~3.3.2.\\

In contrast, the fixed image tracking approach (gray) shows clear deviations, particularly in the systolic phase, where it significantly underestimates pressure compared to traditional FE. This discrepancy arises because the network, constrained by fixed displacements, cannot adjust myocardial motion during training to minimize both the PS pressure control loss (Eq.~\ref{equ:ED/PS}) and the energy balance loss (Eq.~\ref{equ:energy loss2}), leading to poor pressure estimates. This highlights the importance of allowing the optimization of the imaged motions instead of directly adopting them, and that errors in image-derived displacements can lead to inconsistency with biophysics laws, and must be mitigated.\\

The loss component comparisons among IMC-PINN-FE, traditional volume-constrained FE, and fixed image tracking are presented in supplementary S3. The results demonstrate that IMC-PINN-FE consistently achieves low physics losses, comparable to the FE baseline, and substantially lower than those of Fixed Tracking, which has considerable physical inconsistencies in pressure regulation and energy accumulation.\\

The PINN-FE model proposed by Buoso et al.~\cite{buoso2021personalising} also exhibits discrepancies compared to the traditional FE reference P–V loop, particularly in the systolic volumes, which often deviate from the FE references. These deviations arise from the absence of explicit volume constraints to match the imaged volumes, and from the use of an algorithm to construct the isovolumetric phases despite deviations from the measured volumes (i.e., a pressure adjustment scheme that iteratively updates the pressure during the isovolumetric phases toward the target volumes). While this method facilitates the generation of smooth and plausible curves, it introduces artificial vertical truncations in the P–V loop and reduces adaptability to subject-specific variations. In contrast, IMC-PINN-FE simultaneously optimizes both pressure through physics-based constraints, and displacements under volume constraints, enabling a more faithful reproduction of subject-specific P–V dynamics and better alignment with the imaged myocardial motion. However, since smoothing during image motion tracking removed constant volume features from the volume versus time curve, IMC-PINN-FE has the limitation of not explicitly demonstrating the isovolumetric phases.

\subsubsection{Results of Geometric Deformation Accuracy}
\begin{figure}[htbp]
	\centering
	\includegraphics[width=0.7\textwidth]{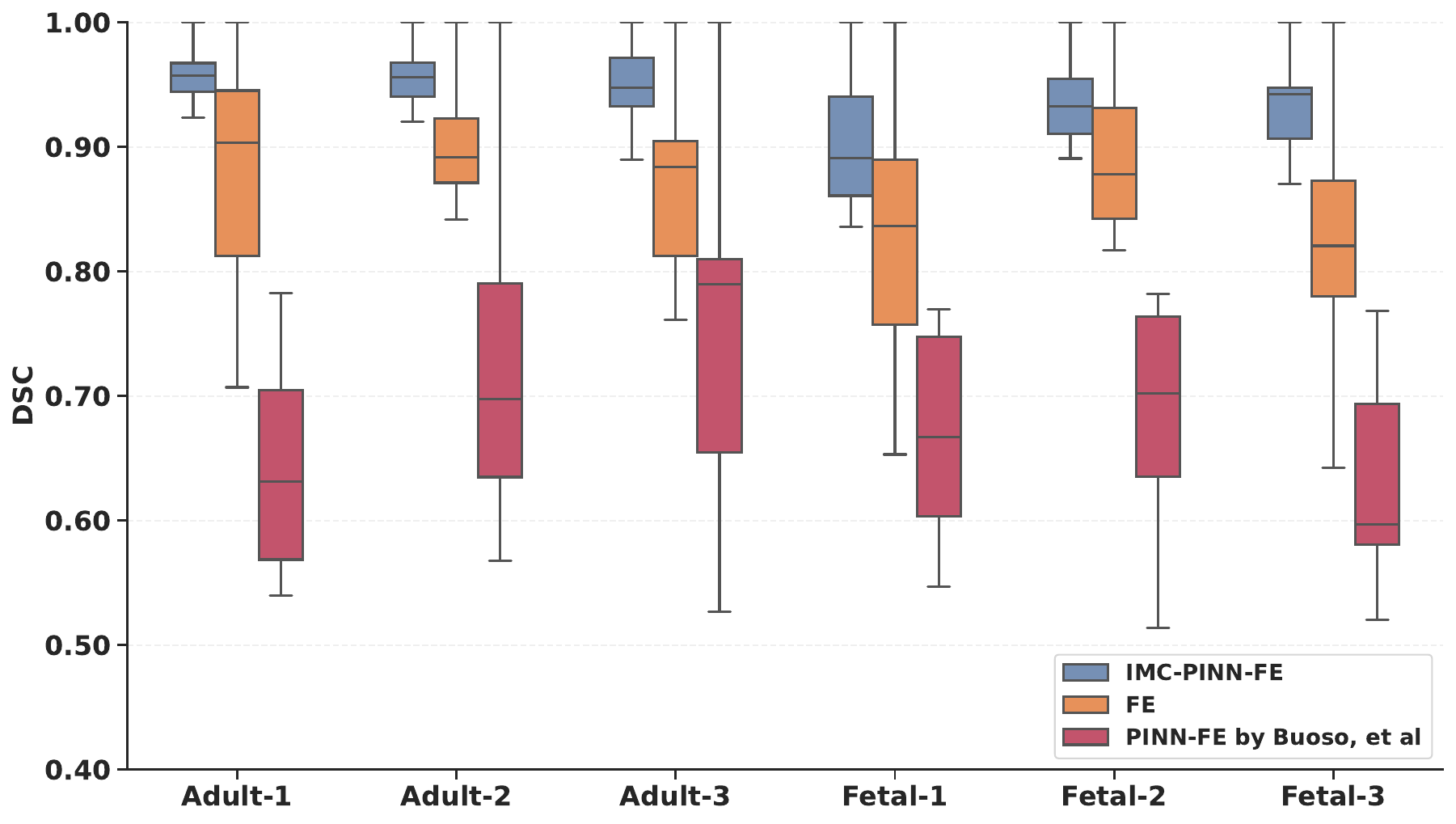}
    \caption{Box plots of DSC evaluation (20 uniformly sampled frames across the cardiac cycle for each of the 6 cases), comparing LV geometries from IMC-PINN-FE, traditional FE, and Buoso et al.’s PINN-FE~\cite{buoso2021personalising} against image-derived results. Orange indicates the DSC between FE outputs and image tracking, blue indicates IMC-PINN-FE vs. image tracking, and red indicates Buoso et al.’s PINN-FE vs. image tracking.}
	\label{FIG:box}
\end{figure}

Fig.~\ref{FIG:box} presents box plots of DSC evaluation over 20 uniformly sampled time frames across the cardiac cycle for each of the 6 cases. DSC quantifies the spatial overlap between the predicted 3D LV shapes (from PINN or FE simulations) and the image-derived ground truth (obtained via segmentation and EFDL motion tracking).\\

Across all cases, IMC-PINN-FE (blue) consistently achieves the highest DSC values (0.927 ± 0.039 across all 6 cases), demonstrating superior shape reconstruction accuracy and robustness. Traditional FE simulations (orange) show moderate performance, with wider spreads and lower medians, indicating greater deviation from image-derived shapes. Buoso et al.’s PINN-FE~\cite{buoso2021personalising} (red) yields the lowest DSC values overall, with large variability and systematic underperformance across both adult and fetal cases. This reflects its limited capability for patient-specific shape recovery, due to reliance on generic shape modes and the absence of image-based constraints.\\

\begin{figure}[htbp]
	\centering
	\includegraphics[width=\textwidth]{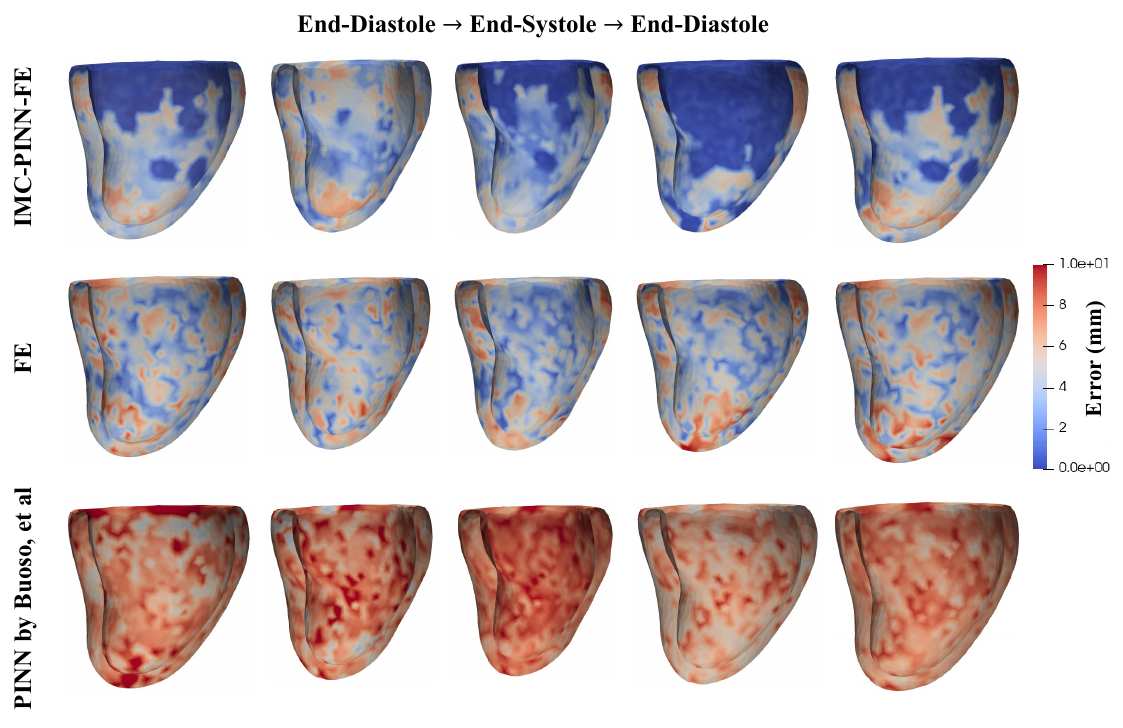}
    \caption{Pointwise deformation error maps across different time points throughout the cardiac cycle (Adult-3). Each row corresponds to a different method: IMC-PINN-FE, traditional FE, and Buoso et al.’s PINN-FE~\cite{buoso2021personalising}. Color encodes the magnitude of pointwise error with respect to the image-derived motion (contour).}
	\label{FIG:Vis_dis}
\end{figure}

These findings are further illustrated in Fig.~\ref{FIG:Vis_dis}, which presents pointwise deformation error maps at five representative time points throughout the cardiac cycle for the Adult-3 case. The color encodes the magnitude of displacement error relative to the image-derived tracking, with red indicating large errors and blue indicating small errors. Traditional FE shows substantial errors, particularly near the apex, where longitudinal motion is overestimated. The method by Buoso et al. exhibits even greater discrepancies from the ground truth, with widespread spatial errors across the myocardium. In contrast, IMC-PINN-FE consistently achieves the lowest deformation errors across the cycle, closely matching the image-derived motion.\\

This discrepancy stems from the lack of constraints to enforce a match of simulated cardiac motions to motions observed in the medical images. The traditional FE modeling generates deterministic solutions based on predefined material properties and boundary conditions, and to enable accurate replication of patient-specific dynamics typically requires extensive manual tuning of such input parameters. As for the PINN method by Buoso et al. \cite{buoso2021personalising}, on top of not enfocing a match with imaged motions, its motion modes are not obtained from individual-specific dynamics, which may contribute further to mismatch with imaged motions. In contrast, IMC-PINN-FE integrates physics-informed constraints with image-derived priors, enabling the model to adapt to personalized motion patterns and produce deformation fields that are both data-consistent and physiologically realistic.

\subsubsection{Comparison of Computational Efficiency}
In addition to its better motion accuracy, IMC-PINN-FE offers significant advantages in computational efficiency. The average runtime for IMC-PINN-FE PINN-FE solver (excluding motion tracking) is 5 minutes and 53 seconds ($\pm$ 193 seconds), while the traditional volume-controlled FE method requires 7 hours and 24 minutes ($\pm$ 15 minutes), for a single cardiac cycle with a 1 ms time step. This corresponds to a nearly 75-fold reduction in computation time, demonstrating the scalability of IMC-PINN-FE for large-scale cardiac simulations.\\

Compared to Buoso et al’s PINN-FE~\cite{buoso2021personalising}, which requires 2 minutes and 20 seconds ($\pm$ 23 seconds) for training for a similar resolution and using the same hardware, IMC-PINN-FE is moderately slower. However, this modest increase in runtime enables direct integration of patient-specific displacement data and volume constraints, which substantially improves the fidelity to imaged cardiac motion. Thus, IMC-PINN-FE is the balanced compromise between accuracy and efficiency.

%% file: Discussion.tex
\subsection{Discussion}

In this study, we introduced IMC-PINN-FE, a novel PINN framework that integrates image motion consistency with biomechanical modeling for patient-specific LV FE simulations. Our method effectively addresses the limitations of traditional FE models and existing PINN-based approaches, which often fail to sufficiently adhere to imaged motion and lack physics-based estimation of myocardial stiffness and contractile function. The proposed framework is applicable to both MRI and echocardiography data. In essence, it enhances the fidelity to imaged motion by incorporating physics constraints to derive more biophysically accurate myocardial dynamics.\\

In our experiments, we further observed that using shape modes for motion encoding improves convergence, whereas unencoded node-wise PINN-FE methods such as \cite{motiwale2024neural} often suffer from convergence issues when applied to patient-specific rather than idealized geometries. The use of shape modes also significantly enhances computational efficiency and reduces training time. For this reason, we adopted a reduced-order, motion-encoded PINN approach.\\

Compared to other reduced-order PINN frameworks, IMC-PINN-FE encodes motion based on patient-specific cardiac dynamics, rather than relying on inter-subject geometric variability from large multi-subject datasets. This strategy may improve patient specificity and eliminates the need for extensive training datasets with ground-truth reconstructions, which are often difficult to obtain in clinical settings.\\

Consequent to the reduced-order modeling, our approach offers substantial computational advantages, reducing simulation time from over 7 hours with traditional iterative FE models to under 10 minutes with IMC-PINN-FE, thereby improving its feasibility for clinical adoption. This efficiency also makes it feasible for IMC-PINN-FE to support high throughput applications, such as building large simulation database for deep learning training, or for large sample clinical testing. \\

In terms of performance, our framework enables robust and efficient training by integrating image-informed constraints with physics-based governing equations. The inference results demonstrate high geometric accuracy (DSC: \(0.927 \pm 0.039\)) for image-derived LV shapes, while maintaining physiological consistency in the predicted P–V loops. Furthermore, the biomechanical parameter estimation component yields reliable personalized predictions of myocardial stiffness and active tension, with relative errors of \(2.81 \pm 1.64\%\) for \(C_{\text{stiffness}}\) and \(2.82 \pm 1.79\%\) for \(T_{\text{max}}\).\\

Although IMC-PINN-FE uses the same equation systems as traditional FE modeling, it yields different results with a better match to imaged motion, and yet produces a very similar P-V loop. This apparent discrepancy can be explained by the range of solution that are possible to the FE governing equations, and the difficult in establishing a uniqueness of solution with the large degree of freedom. Essentially, IMC-PINN-FE is able to find another displacement solution that is close to the imaged motions but which still conforms to the physics equations. The key to enabling this is to not directly constrain IMC-PINN-FE to the imaged displacement fields, in line with the understanding that motion extraction from images has errors, but to allow it to have some deviations and space to finds the closest solution conforming to both motion and physics constraints. Overall, IMC-PINN-FE offers an accurate and data-efficient alternative to conventional FE methods, which are typically time-consuming.

\subsection{Limitations and Future Work}

Despite the promising performance of IMC-PINN-FE, several limitations remain. A primary limitation is the framework’s dependency on image quality and tracking accuracy. Since the motion modes in IMC-PINN-FE are derived from image tracking, inaccuracies or noise in the tracking process can directly affect the reliability of the estimated myocardial motion. This issue warrants further investigation. In addition, incorporating uncertainty quantification techniques could help assess the confidence of predictions under varying image qualities and tracking conditions, thereby improving model robustness.\\

Another current limitation is the lack of integration with a Windkessel model, although this can be readily addressed in future work. Moreover, the process of selecting temporal resolution could be refined through adaptive sampling strategies, with the goal of optimizing computational efficiency while preserving accuracy. In the current implementation, frames are uniformly sampled at a fixed interval of 1~ms, but this setting can be adjusted in future work to focus sampling on periods of high motion complexity or uncertainty.